\documentclass[11pt,a4paper]{article}

\usepackage{amsmath,amssymb}
\usepackage{epsfig,graphicx}
\usepackage{graphicx}
\usepackage{rotating}
\usepackage{cancel}
\usepackage{bm}
\usepackage{xcolor}
\usepackage{comment}
\usepackage{cite}
\usepackage{psfrag}

\usepackage{caption}
\usepackage{subcaption}
\usepackage{enumerate}
\graphicspath{{img/}}

\renewcommand\({\left(}
\renewcommand\){\right)}
\renewcommand\[{\left[}

\newcommand{\exclude}[1]{}

\def\bra{\langle}
\def\ket{\rangle}
\def\beq{\begin{equation}}
\def\eeq{\end{equation}}

\newcommand{\amp}[2][0]{\langle #2 \lvert \hat{x} \rvert #1 \rangle}
\newcommand{\kamp}[3]{\langle #1 \lvert \hat{x}^{#2} \rvert #3 \rangle}

\begin{document}
\numberwithin{equation}{section}
\title{
\vspace{2.5cm} 
\Large{\textbf{Exploring High Multiplicity Amplitudes\\ in Quantum Mechanics
\vspace{0.5cm}}}}

\author{Joerg Jaeckel and Sebastian Schenk\\[2ex]
\small{\em Institut f\"ur theoretische Physik, Universit\"at Heidelberg,} \\
\small{\em Philosophenweg 16, 69120 Heidelberg, Germany}\\[0.8ex]}

\date{}
\maketitle

\begin{abstract}
\noindent
Calculations of $1\to N$ amplitudes in scalar field theories at very high multiplicities exhibit an extremely rapid growth with the number $N$ of final state particles. This either indicates an end of perturbative behaviour, or possibly even a breakdown of the theory itself. It has recently been proposed that in the Standard Model this could even lead to a solution of the hierarchy problem in the form of a ``Higgsplosion''~\cite{Khoze:2017tjt}. 
To shed light on this question we consider the quantum mechanical analogue of the scattering amplitude for $N$ particle production in $\phi^4$ scalar quantum field theory, which corresponds to transitions $\amp{N}$ in the anharmonic oscillator with quartic coupling $\lambda$.
We use recursion relations to calculate the $\amp{N}$ amplitudes to high order in perturbation theory.
Using this we provide evidence that the amplitude can be written as $\amp{N}\sim \exp(F(\lambda N)/\lambda)$ in the limit of large $N$ and $\lambda N$ fixed. We go beyond the leading order and provide a systematic expansion in powers of $1/N$.
We then resum the perturbative results and investigate the behaviour of the amplitude in the region where tree-level perturbation theory violates unitarity constraints. The resummed amplitudes are in line with unitarity as well as stronger constraints derived by Bachas~\cite{Bachas:1991fd}.
We generalize our result to arbitrary states and powers of local operators $\kamp{N}{q}{M}$ and confirm that, to exponential accuracy, amplitudes in the large $N$ limit are independent of the explicit form of the local operator, i.e.~in our case $q$.
\end{abstract}

\newpage

\section{Introduction}

Multiparticle amplitudes in weakly coupled scalar quantum field theories have been attracting quite some interest in the past, since they seem to exhibit factorial growth with the number of particles produced~\cite{Cornwall:1990hh, Goldberg:1990qk, Brown:1992ay, Voloshin:1992mz, Argyres:1992np, Smith:1992kz, Smith:1992rq}.
With the discovery of a scalar Higgs boson~\cite{Aad:2012tfa,Chatrchyan:2012xdj} this has turned into a problem of the Standard Model, with an explicit maximal scale where either novel non-perturbative behaviour or new physics must set in~\cite{Voloshin:1992rr, Jaeckel:2014lya}. The relevant scale can be estimated to be $\lesssim1600~{\rm TeV}$ at tree-level, but higher order corrections indicate it could even be much lower, possibly even within the range of the next generation of colliders.

A particularly interesting form of non-perturbative behaviour could be a ``Higgsplosion'' and ``Higgspersion'' effect proposed in~\cite{Khoze:2017tjt}. Here, the increase in the $1\to N$ amplitudes leads to a rapid growth in the decay width of the particle. The large width suppresses the propagator, effectively cutting of loop-integrals at the scale where $1\to N$ amplitudes become large. Since this scale is low, this can address the hierarchy problem. At the same time the low scale provides for a potentially rich phenomenology~\cite{Khoze:2017lft, Khoze:2017uga, Khoze:2018bwa}.

Evidence for the rapid growth of $1\to N$ amplitudes arises from perturbative~\cite{Cornwall:1990hh, Goldberg:1990qk, Brown:1992ay, Voloshin:1992mz, Argyres:1992np, Smith:1992kz, Smith:1992rq} as well as semiclassical~\cite{Son:1995wz,Khoze:2017ifq} calculations.
An interesting and useful limit to consider is $N\to\infty$, $\lambda N=const$. In this double scaling limit the amplitudes
can be written in an exponential form~\cite{Voloshin:1992nu,Khlebnikov:1992af,Libanov:1994ug,Libanov:1995gh,Bezrukov:1995qh,Libanov:1996vq}
\begin{equation}
	\mathcal{A} \left( \phi^{\star} \to N \phi \right)=\langle N|\phi|0\rangle \sim  \exp \left( \frac{1}{\lambda} F \left( \lambda N \right) \right).
\end{equation}
Here the function $F$ is a function of the combination $\lambda N$ only. 

In the semiclassical\footnote{These techniques for QFT are modelled after the Landau method in quantum mechanics~\cite{Landau:1991wop}.} approach~\cite{Khlebnikov:1992af,Son:1995wz,Khoze:2017ifq} this form is inherent and indeed it only attempts to compute the function $F$.
The semiclassical calculations also rely on the assumption that, to exponential accuracy, the amplitude in question is independent of the precise form of the local operator~\cite{Son:1995wz,Khoze:2017ifq,Landau:1991wop}\footnote{We also briefly mention at the end of Section~\ref{sec:arbitrarylocal} that this statement is not sufficient to guarantee that no exponential corrections arise, if the calculated operator is itself an exponential of the field operator.}, e.g. 
\begin{equation}
\langle N|\phi|0\rangle\sim \langle N|\phi^2 |0\rangle.
\end{equation}

It is one of the main aims of the present work to support these statements with explicit calculations.

\bigskip

Phenomenologically the sign of the function $F$ is crucial. If $F>0$ for some value $\lambda N> 0$ we can always find a rapid growth of the amplitudes in a limit where we keep $\lambda N$ fixed at this value and send $\lambda\to 0$.
We are therefore particularly interested in establishing the sign of the function $F$. 
In the symmetric $\phi^4$-theory the tree-level calculation results in a positive $F$ at $\lambda N>8e$. But the next correction is negative~\cite{Voloshin:1992nu,Libanov:1997nt}, leaving the case unclear but providing hope for convergence.
In the spontaneously broken system the situation is more dire.
The loop correction is also positive~\cite{Smith:1992rq}. This is also supported in the semiclassical calculation~\cite{Khoze:2017ifq} which also yields a positive sign at large $\lambda N$.
In this paper we will focus on the simpler symmetric case. 

Although we are ultimately interested in quantum field theory we will consider here the quantum mechanical analogue of the symmetric $\phi^4$-theory, i.e.~the anharmonic oscillator with a quartic potential
\begin{equation}
V(x) = x^2 + \lambda x^4.
\end{equation}
This will allow us to do explicit calculations to high orders and perform resummations that enable us to investigate the large $\lambda N$ behaviour. While this can give us important insights into the desired high multiplicity amplitudes we should nevertheless be aware that quantum field theory in four dimensions provides for additional structure and complications, e.g.~a non-trivial phase space.

Quantum mechanics has already been a testbed for many investigations of high order perturbation theory (see, e.g.~\cite{Bender:1969si, Bender:1990pd,Loeffel:1970fe,ZinnJustin:2004ib, Dunne:2014bca, Sulejmanpasic:2016fwr, Serone:2017nmd}). However, with few exceptions~\cite{Bachas:1991fd,Bachas:1992dw,Brown:1992ay,Cornwall:1991gx,Cornwall:1992ip,Cornwall:1993rh,Khlebnikov:1994dc}, most of these focussed on energy levels and wave functions in or near the ground state. In contrast here we consider transition amplitudes to highly excited states.

This work is structured as follows.
Section \ref{sec:perturbation_theory} reviews the general method of reconstructing the wave functions of the anharmonic oscillator systematically by exploiting recursive methods.
These are then used to derive a perturbative expression for the multiparticle amplitudes in Section \ref{sec:amplitudes}.
Section \ref{sec:holy_grail} is devoted to investigating the exponentiation of the multiparticle amplitude in the regime $N \to \infty$ and $\lambda N$ fixed.
In particular, the function $F$ is computed explicitly to a high perturbative order.
We also consider and compute corrections to the exponent that are suppressed by powers of $1/N$.
We then use Pad\'e techniques to resum the perturbative series and show that it remains negative and avoids problems with unitarity at the point where unitarity breaks down at tree-level.
We also compare to the upper bounds derived in~\cite{Bachas:1991fd}.
In Section~\ref{sec:arbitrarylocal} we then extend our results to amplitudes involving more general local (in time) operators and show that they indeed only differ in the pre-exponential factor as assumed in semiclassical calculations.
Finally, we briefly summarize and conclude in Section~\ref{sec:conclusions}.

\section{Reconstructing Wave Functions of the Anharmonic Oscillator}
\label{sec:perturbation_theory}
For high order calculations recursion relations are an efficient way to organize perturbation theory.
Let us briefly recall the methods developed by Bender and Wu~\cite{Bender:1969si, Bender:1990pd}, whose derivation we follow closely.

We are interested in finding solutions to the Schroedinger equation
\begin{equation}
\left( -\frac{d^2}{dx^2} + V(x) - E \right) \psi = 0
\label{eq:schroedinger}
\end{equation}
for the anharmonic oscillator potential with a unique global minimum
\begin{equation}
V(x) = x^2 + \lambda x^4 \; , \quad \lambda > 0 \, .
\label{eq:potential}
\end{equation}

In order to find the $N$-th energy level $E_N$ and its corresponding eigenfunction $\psi_N$ we make use of the recursive methods developed in~\cite{Bender:1969si, Bender:1990pd}. We start by introducing the polynomial ansatz
\begin{equation}
\psi_N (x, \lambda) = c_N e^{- \frac{x^2}{2}} \sum_{n=0}^\infty \lambda^n B_n^N (x)
\label{eq:ansatz_wavefunction}
\end{equation}
where $c_N$ is a normalization constant and the functions $B_n^N(x)$ are polynomials of the form
\begin{equation}
B_n^N (x) = \sum_k B_{n,k}^N x^k \, .
\end{equation}
In a similar manner we make an ansatz for the energy levels by writing
\begin{equation}
E_N (\lambda) = 2N+1 + \sum_{n=1}^\infty \lambda^n a_n^N \, .
\label{eq:ansatz_energy}
\end{equation}
For simplicity of notation we will drop the index $N$ of the $N$-th level from now on and keep it in mind implicitly.
Plugging this ansatz into the Schroedinger equation \eqref{eq:schroedinger}, one can derive a recursion relation for the polynomial functions $B_n(x)$ as the coefficient of the perturbative series at order $\mathcal{O} \left( \lambda^n \right)$, which is given by
\begin{equation}
2x B_n^\prime - B_n^{\prime \prime} + x^4 B_{n-1} = 2N B_n + \sum_{k=0}^{n-1} a_{n-k} B_k
\label{eq:B_recursion}
\end{equation}
where $B_n^\prime = \frac{d}{dx} B_n(x)$. 

At leading order $n=0$ we obtain the differential equation
\begin{equation}
B_0^{\prime \prime} - 2x B_0^\prime + 2N B_0 = 0
\end{equation}
which is exactly solved by the Hermite polynomials of order $N$, i.e.~$B_0^N (x) = H_N(x)$.

We now want to go beyond leading order.
Since we assume the functions $B_n(x)$ to be polynomials, the relation~\eqref{eq:B_recursion} can be translated into a similar recursive relation for their polynomial coefficients $B_{n,k}$.
It takes the form
\begin{equation}
(k+1)(k+2) B_{n,k+2} = 2 (k - N) B_{n,k} + B_{n-1, k-4} - \sum_{p=0}^{n-1} a_{n-p} B_{p, k} \, .
\label{eq:Bcoeff_recursion}
\end{equation}

In order to find the tower of coefficients $B_{n,k}$ we again have to solve the recursive relation \eqref{eq:Bcoeff_recursion} by distinguishing between two different physical cases.
We can either consider even or odd $N$, i.e.~even or odd wavefunctions $\psi_N$ with respect to parity transformations.

\begin{enumerate}[(i)]
\item For even $N$ we infer that all the odd parts of the polynomial expansion in $B_n(x)$ should vanish, i.e.~$B_{n,k} = 0$ for $k \in 2 \mathbb{N} + 1$.
In addition we can evaluate \eqref{eq:Bcoeff_recursion} for $k=0$ to arrive at
\begin{equation}
2 B_{n,2} + 2N B_{n,0} = - \sum_{p=0}^{n-1} a_{n-p} B_{p,0}\, .
\end{equation}
If we choose\footnote{Since we are essentially solving a differential equation, we are allowed to fix the boundary conditions.} $B_{n,0} = 0$ for $n \geq 1$, we can immediately read off the values for energy expansion coefficients, which are 
\begin{equation}
a_n = -2 \frac{B_{n,2}}{B_{0,0}} \quad \forall n \geq 1\, .
\end{equation}

\item Similarly to the considerations for even $N$ we can derive a relation for the $a_n$ when $N$ is an odd number.
In this case we infer that all the even parts of the polynomial expansion in $B_n(x)$ should vanish, i.e.~$B_{n,k} = 0$ for $k \in 2 \mathbb{N}$.
Evaluating \eqref{eq:Bcoeff_recursion} for $k=1$ yields
\begin{equation}
6 B_{n,3} - 2(1-N) B_{n,1} = - \sum_{p=0}^{n-1} a_{n-p} B_{p,1} \, .
\end{equation}
Together with the choice $B_{n,1} = 0$ for $n \geq 1$ we arrive at the final expression for the energy expansion coefficients
\begin{equation}
a_n = -6 \frac{B_{n,3}}{B_{0,1}} \quad \forall n \geq 1 \, .
\end{equation}
\end{enumerate}

We can summarize both cases in the following relation
\begin{equation}
(k+1)(k+2) B_{n,k+2} - 2(k-N) B_{n,k} - B_{n-1,k-4} = \begin{cases} \hfill \frac{2}{B_{0,0}} \sum_{p=0}^{n-1} B_{n-p,2} B_{p,k}   \hfill & N \mathrm{even} \\
      \hfill \frac{6}{B_{0,1}} \sum_{p=0}^{n-1} B_{n-p,3} B_{p,k} \hfill & N \mathrm{odd} \end{cases} \, .
\label{eq:Bcoeff_recursion_expl}
\end{equation}
Keeping $n$ fixed this relation can easily be solved for the full $k$-tower of coefficients $B_{n,k}$, where we use $B_0^N (x) = H_N(x)$.

The recursion relation becomes vacuous for sufficiently large values of $k$, i.e.~the coefficients become proportional to each other $B_{n,k+2} \propto B_{n,k}$ for $k \geq k_{\mathrm{max}}$.
An explicit computation shows that this threshold is given by $k_{\mathrm{max}} = N + 4n + 2$.
From a physical point of view this is consistent with the fact that the wave function has to be square-integrable.
To achieve this we use the proportionality to truncate the polynomial expansion at order $k_{\mathrm{max}}$, i.e.~$B_{n,k} = 0$ for $k \geq N + 4n + 2$.

In fact, this is all the information we need to reconstruct the $N$-th level wavefunction $\psi_N$ up to arbitrary perturbative order $\mathcal{O} \left( \lambda^n \right)$.
For instance, the first two levels read
\begin{align}
\psi_0 (x) &= c_0 e^{-\frac{x^2}{2}} \left[1 - \frac{\lambda}{8} {\left(x^{4} + 3 x^{2}\right)} + \frac{\lambda^2}{384} {\left(3 x^{8} + 26 x^{6} + 93 x^{4} + 252 x^{2}\right)} + \mathcal{O} \left( \lambda^3 \right) \right] \\
\psi_1 (x) &= c_1 e^{-\frac{x^2}{2}} \left[2x- \frac{\lambda}{4} {\left(x^{5} + 5 x^{3}\right)} + \frac{\lambda^2}{192} \, {\left(3 \, x^{9} + 38 \, x^{7} + 177 \, x^{5} + 660 \, x^{3}\right)} + \mathcal{O} \left( \lambda^3 \right) \right].
\end{align}

Finally, we need to determine the wave function normalization $c_N$.
It is given by the condition,
\begin{equation}
\bra N | N \ket = \int_\mathbb{R} dx \; \bar{\psi}_N \psi_N = \lvert c_N \rvert^2 \int_\mathbb{R} dx \; e^{-x^2} \sum_n \lambda^n B_n^N(x) \sum_{p} \lambda^p B_p^N(x) \overset{!}{=} 1
\label{eq:norm_int}
\end{equation}
which again is a perturbative series in powers of the coupling of the theory.
At leading order we recover the well known result
\begin{equation}
\lvert c_N \rvert_0^2 = \frac{1}{\sqrt{\pi} 2^N N!} \, .
\label{eq:norm_tree}
\end{equation}

\section{Vacuum Transition Amplitudes}
\label{sec:amplitudes}

In the previous section we recalled how to reconstruct the $N$-th level wave function $\psi_N$ up to arbitrary perturbative order in the coupling from a tower of recursive relations.
We can use these results to compute transition amplitudes in the anharmonic oscillator of the form
\begin{equation}
\bra N | \hat{x} | 0 \ket = \int_\mathbb{R} dx \; \bar{\psi}_N x \psi_0 = \int_\mathbb{R} dx \; x e^{-x^2} \sum_n \lambda^n B_n^N(x) \sum_{p} \lambda^p B_p^0(x) \, .
\end{equation}
Since the theory enjoys a $\mathbb{Z}_2$ symmetry at the Lagrangian level (i.e.~parity in our case), we can immediately conclude that
\begin{equation}
\bra N | \hat{x} | 0 \ket = 0 \quad \forall N \in 2 \mathbb{N} \, .
\label{eq:amplitude_even}
\end{equation}
Using a Feynman diagram picture it is straightforward to see that the tree-level contribution of the amplitude is expected to be
\begin{equation}
\bra N | \hat{x} | 0 \ket_{\mathrm{tree}} \sim \lambda^{\frac{N-1}{2}}.
\label{eq:treelevel_naive}
\end{equation}

Plugging the polynomial coefficients derived from \eqref{eq:Bcoeff_recursion} into the explicit expression for $\bra N | \hat{x} | 0 \ket$ yields\footnote{Note carefully that in this manipulation we are exchanging the summation and integration even though it might not be strictly allowed in this case. Indeed this is a point where problems of perturbation theory may arise.}
\begin{equation}
\bra N | \hat{x} | 0 \ket = \sum_{n=0}^\infty \lambda^n t_n^N \enspace \mathrm{with} \enspace t_n^N = \sum_{p=0}^n \sum_{k=0}^{N+4p} \sum_{l=0}^{4(n-p)} B_{p, k}^N B_{n-p,l}^0 \Gamma \( \frac{k+l+2}{2} \) \, .
\label{eq:amplitude_recursion}
\end{equation}
This form of the amplitude is in perfect agreement with the naive expectation \eqref{eq:treelevel_naive}, because we obtain the non-trivial relation  $t_n^N \equiv 0$ for $n < (N-1)/2$.

Since the computation of $\bra N | \hat{x} | 0 \ket$ is completely similar to the computation of $\bra N | N \ket$ defined in \eqref{eq:norm_int}, we can also immediately give an expression for the wave function normalization,
\begin{equation}
\bra N | N \ket = \sum_{n=0}^\infty \lambda^n m_n^N \enspace \mathrm{with} \enspace m_n^N = \sum_{p=0}^n \sum_{k=0}^{N+4p} \sum_{l=0}^{N+4(n-p)} B_{p, k}^N B_{n-p,l}^N \Gamma \( \frac{k+l+1}{2} \) \, .
\label{eq:norm_recursion}
\end{equation}

In principle, the amplitude \eqref{eq:amplitude_recursion} together with the recursive relation of the polynomial coefficients \eqref{eq:Bcoeff_recursion} is sufficient to compute $\bra N | \hat{x} | 0 \ket$ up to arbitrary order in $\lambda$.
However, we are interested in the analytic $N$-dependence of $\bra N | \hat{x} | 0 \ket$, which is contained in \eqref{eq:Bcoeff_recursion} and \eqref{eq:amplitude_recursion} only implicitly.
Consequently, we need to match assumptions on the analytic behaviour to the numerical expression.

In particular, we observe that $\bra N | \hat{x} | 0 \ket$ is of polynomial form
\begin{equation}
\bra N | \hat{x} | 0 \ket = \bra N | \hat{x} | 0 \ket_{\mathrm{tree}} \sum_{k=0}^\infty \lambda^k P_{2k} \left( N \right)
\end{equation}
where $P_k \left( N \right)$ denotes a polynomial of degree $k$ with coefficients in $N$.
Since we can compute $\bra N | \hat{x} | 0 \ket$ for any $N$ recursively, we can determine these coefficients of $P_k$ separately in order to completely fix the analytic form of $\bra N | \hat{x} | 0 \ket$.
For instance, an explicit computation yields
\begin{eqnarray}
\label{eq:amplitude_explicit}
&&\!\!\!\!\!\!\!\!\bra N | \hat{x} | 0 \ket
\\\nonumber
 &&= \bra N | \hat{x} | 0 \ket_{\mathrm{tree}} \left\{ 1 - \lambda \frac{N}{16} \left( 17N+20 \right) + \lambda^2 \frac{N}{512} \left( 289N^3 + 1680N^2 + 2072N + 2060 \right) + \mathcal{O}(\lambda^3) \right\}
\end{eqnarray}
where the tree-level contribution is given by
\begin{equation}
\bra N | \hat{x} | 0 \ket_{\mathrm{tree}} = \sqrt{\pi} N! \left( \frac{\lambda}{4} \right)^{\frac{N-1}{2}}.
\label{eq:tree_amplitude}
\end{equation}
In general the explicit form of $\bra N | \hat{x} | 0 \ket$ in \eqref{eq:amplitude_explicit} can be computed to arbitrary order in $\lambda$ with correspondingly long expressions for each coefficient.

Again, a completely similar computation can be done for the wave function normalization to arbitrary order in the coupling.
It is given by
\begin{equation}
\bra N | N \ket = \bra N | N \ket_0 \left\{ 1 - \lambda \frac{15}{16} \left(2 N + 1\right) + \lambda^2 \frac{1}{512} \left( 863 \left(2N+1\right)^2 + 718 \right) + \mathcal{O} \left( \lambda^3 \right) \right\},
\label{eq:norm_explicit}
\end{equation}
where $\bra N | N \ket_0 = \sqrt{\pi} 2^N N!$ was computed before in \eqref{eq:norm_tree}.
It is remarkable that the leading terms of the normalization are an expansion in powers of $\lambda N$ in contrast to the $\lambda N^2$ asymptotics of the amplitude.
This implies that the leading order behaviour in $\lambda N^2$ of $\bra N | \hat{x} | 0 \ket$ will not be affected by the normalization.
Nevertheless, since only normalized wave functions and transition amplitudes are physical, we will focus on computing the quantity,
\begin{equation}
\mathcal{A}_N \equiv \frac{\bra N | \hat{x} | 0 \ket}{\sqrt{\bra N | N \ket} \sqrt{\bra 0 | 0 \ket}} \, .
\label{eq:amplitude_norm_independent}
\end{equation}
In fact, the normalized amplitude is of the form
\begin{eqnarray}
\label{eq:amplitude_normalized}
&&\!\!\!\!\!\!\!\!
\mathcal{A}_N = \mathcal{A}_N^{\mathrm{tree}}
\\\nonumber
&&\,\,\,\times \left\{1 - \frac{\lambda}{16} \left( 17N^2 + 5N - 12 \right) + \frac{\lambda^2}{512} \left( 289N^4 + 1170N^3 + 13N^2 +664N - 944 \right) + \mathcal{O} \left( \lambda^3 \right) \right\}
\end{eqnarray}
where the tree-level factor is given by
\begin{equation}
\mathcal{A}_N^{\mathrm{tree}} = \frac{1}{\sqrt{2}} \sqrt{N!} \left( \frac{\lambda}{8} \right)^{\frac{N-1}{2}} \, .
\label{eq:amplitude_normalized_tree}
\end{equation}

Note that in quantum mechanics the reduction of the prefactor to $\sqrt{N!}$ instead of $N!$ arises from the normalization condition. In quantum field theory a similar role is played by the phase space integration of the squared matrix element which contains a factor of $1/N!$ effectively reducing the growth by a factor $\sqrt{N!}$.

\section{Exponentiation of the Amplitude and the Holy Grail Function}
\label{sec:holy_grail}
As already mentioned in the introduction many considerations for high multiplicity amplitudes are based on its exponential form.
Let us briefly consider what is special about this. In principle we can write any function $B(\lambda,N)$ as,
\begin{equation}
B(\lambda,N)=\exp\left[\log(B(\lambda,N) \right]=\exp\left[L(\lambda,N)\right] \, .
\end{equation}
We now take the limit $N\to \infty$, $\lambda N=const$. Assuming that $L$ behaves as 
\begin{equation}
\label{assumption}
L\sim N^{\kappa}+{\mathcal{O}}(1/N)
\end{equation}
at large $N$ we have,
\begin{equation}
L(\lambda,N)=N^{\kappa}\hat{L}(\lambda N)+{\mathcal{O}}\left(\frac{1}{N}\right)=\frac{1}{\lambda^{\kappa}}(\lambda N)^{\kappa}\hat{L}(\lambda N)+{\mathcal{O}}\left(\frac{1}{N}\right),
\end{equation}
where the function $\hat{L}$ now only depends on the combination $\lambda N$.
Defining
\begin{equation}
f(\lambda N)=(\lambda N)^{\kappa}\hat{L}(\lambda N),
\end{equation}
we arrive at the desired form,
\begin{equation}
B(\lambda,N)\sim\exp\left(\frac{f(\lambda N)}{\lambda^{\kappa}}\right).
\end{equation}
This statement can be generalized to also include $1/N$ corrections. Indeed we can write,
\begin{equation}
\label{higher}
B(\lambda,N)\sim\exp\left(\frac{1}{\lambda^{\kappa}}\left[f_{0}(\lambda N)+\frac{1}{N}f_{1}(\lambda N)+\ldots\right]\right).
\end{equation}
Below we will also explicitly compute those $1/N$ corrections for the case of the anharmonic oscillator.

So far this is a rather general statement. However, the assumption~\eqref{assumption} is crucial. Indeed, to obtain~\eqref{higher} we also need to be able to expand in powers of $1/N$.
In perturbation theory we will find below that this assumption is fully justified, $\kappa=1$ and going to sufficiently high order {\emph{all}} coefficients in perturbation theory can be recovered, not only those that are dominant in the limit $N\to \infty$, $\lambda N=const$.

Let us illustrate that this is far from trivial by considering the following example,
\begin{equation}
B(\lambda, N)=2\cosh\left(\frac{1}{\lambda}(\lambda N)^2\right)=2+\lambda^2N^{4}
+\frac{1}{12}\lambda^4N^{8}+\ldots .
\end{equation}
In the considered limit one quickly finds,
\begin{equation}
f(\lambda N)=(\lambda N)^2
\end{equation}
and this remains true to all orders in $1/N$.
However, the reconstructed function
\begin{equation}
B(\lambda,N)\sim \exp\left(\frac{f(\lambda N)}{\lambda}\right)=1+\lambda N^2+\frac{1}{2}\lambda^2N^{4}+\frac{1}{6}\lambda^3N^{6}+\frac{1}{24}\lambda^4N^{8}+\ldots,
\end{equation}
contains coefficients not in the expansion of the original function.
It is straightforward to check that the logarithm of the original function differs from $f(\lambda N)$ by,
\begin{equation}
\log(B(\lambda,N))-\frac{f(\lambda N)}{\lambda}=\exp\left(-2(\lambda N)N\right)-\frac{1}{2}\exp\left(-4(\lambda N)N \right)+\ldots .
\end{equation}
While the assumption Eq.~\eqref{assumption} is fulfilled, the difference is that the true logarithm of the original function contains a part that is exponentially suppressed as $1/N\to 0$ and $\lambda N=const$ (i.e. $N\to\infty$). 
This part, that is not a function of just the combination $\lambda N$, contains the information about all the coefficients that are not correctly reproduced by the exponential $\exp(f(\lambda N)/\lambda)$.

In the case of the anharmonic oscillator we find below the remarkable property that we have exact exponentiation in the sense that all coefficients of perturbation theory can be recovered from the exponent, if it is calculated to sufficiently high order.\footnote{In more general quantum mechanical systems this is not generally true and the statement has to be modified, as we will discuss in future work~\cite{inpreparation}.}

\bigskip

In the last section we demonstrated a procedure how to compute the amplitude $\mathcal{A}_N$, given in \eqref{eq:amplitude_norm_independent}. It was shown to factorize into $\mathcal{A}_N = \mathcal{A}_N^{\mathrm{tree}} \mathcal{A}_\Sigma$, where for simplicity of notation we use (cf.~\eqref{eq:amplitude_normalized})
\begin{equation}
\mathcal{A}_\Sigma = 1 - \frac{\lambda}{16} \left( 17N^2 + 5N - 12 \right) + \frac{\lambda^2}{512} \left( 289N^4 + 1170N^3 + 13N^2 +664N - 944 \right) + \mathcal{O} \left( \lambda^3 \right) \, .
\label{eq:amplitude_factorize}
\end{equation}

From the explicit form of $\mathcal{A}_\Sigma$ an intriguing observation can be made.
The first few leading terms of the amplitude in $\lambda N^2$
\begin{equation}
\mathcal{A}_\Sigma \sim 1 - \frac{17}{16} \lambda N^2 + \frac{289}{512} \lambda^2 N^4 - \frac{4913}{24576} \lambda^3 N^6 + \mathcal{O} \( \lambda^4 N^8 \)
\end{equation}
can in fact be written as an exponential function which takes the form
\begin{equation}
\mathcal{A}_\Sigma \sim \exp \( -\frac{17}{16} \lambda N^2 \) \, .
\end{equation}
This crucial property of $\mathcal{A}_\Sigma$ supports the conjecture that in the double scaling limit $N \to \infty$ and $\lambda \to 0$ with $\lambda N$ fixed the full amplitude can be written in exponential form \cite{Voloshin:1992nu, Libanov:1994ug, Libanov:1995gh, Bezrukov:1995qh, Libanov:1996vq, Son:1995wz}
\begin{equation}
\mathcal{A}_N \sim 
\exp \( \frac{1}{\lambda} F \( \lambda N \) \) \, ,
\label{eq:amplitude_exp_general}
\end{equation}
where $F$ is sometimes called \textit{holy grail function}. Note that at this point we neglect corrections of order $1/N$ to it which we will discuss below.

In order to obtain \eqref{eq:amplitude_exp_general} we write
\begin{equation}
\mathcal{A}_N = \mathcal{A}_N^{\mathrm{tree}} \mathcal{A}_\Sigma \sim \exp \left\{ \frac{1}{\lambda} \left( F^{\mathrm{tree}} + F_\Sigma \right) \right\} \, .
\end{equation}

That is, $F$ can be separated into a tree-level and a higher order contribution
\begin{equation}
F \left( \lambda N \right) = F^{\mathrm{tree}} \left( \lambda N \right) + F_\Sigma \left( \lambda N \right)\, .
\end{equation}
$F^{\mathrm{tree}}$ and $F_\Sigma$ correspond to $\mathcal{A}_N^{\mathrm{tree}}$ and $\mathcal{A}_\Sigma$ of the full amplitude, respectively.

\begin{figure}[t]
\centering
\includegraphics[width=0.5\textwidth]{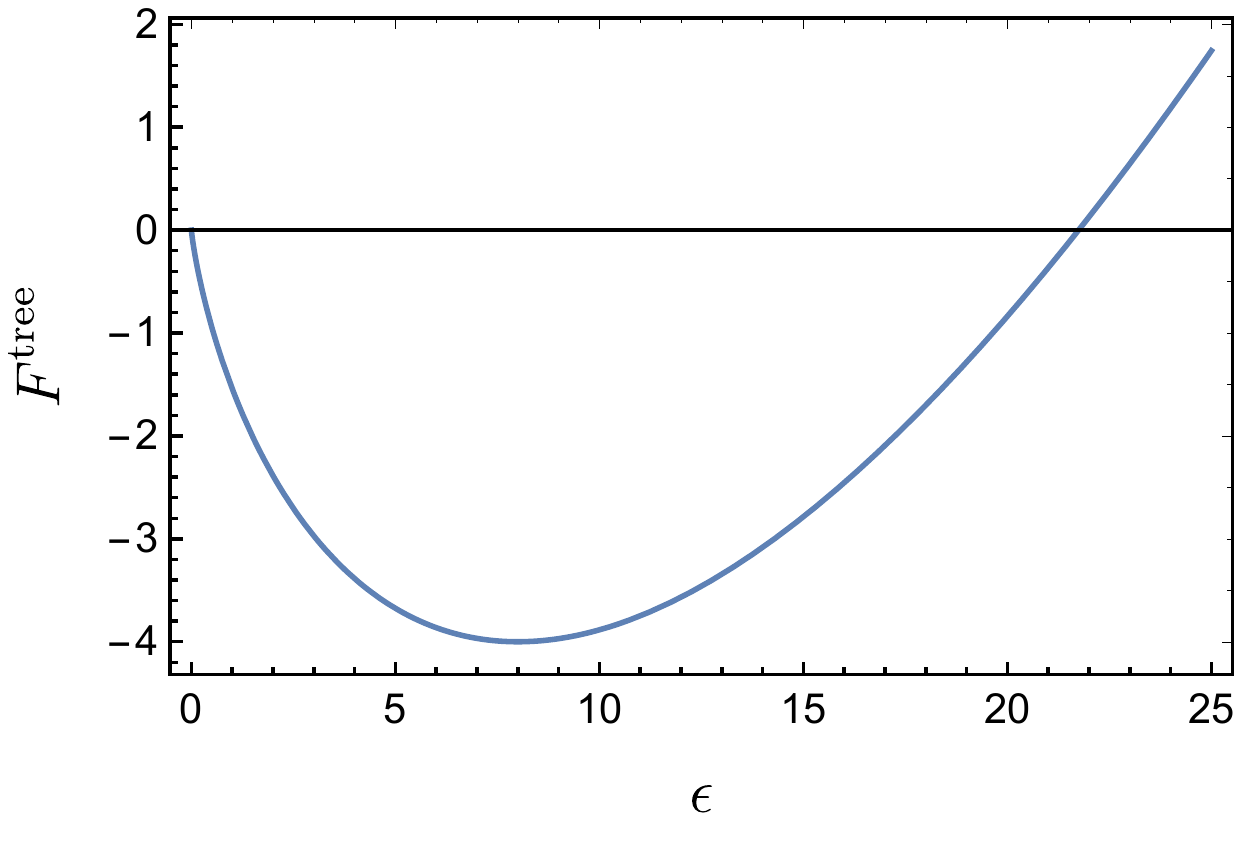}
\caption{Holy grail function $F^{\mathrm{tree}}$ corresponding to the tree-level amplitude $\mathcal{A}_N^{\mathrm{tree}}$ in the double scaling limit $N\to\infty$, $\epsilon=\lambda N=const$ (corrections of order $1/N$ are neglected). It exhibits a global minimum at $\epsilon_{\mathrm{min}} = 8$ and root at $\epsilon_0 = 8e$.}
\label{fig:F_tree}
\end{figure}

For convenience we use in the following the abbreviation,
\begin{equation}
\epsilon=\lambda N \, .
\end{equation}
It is straightforward to write the tree-level amplitude \eqref{eq:amplitude_normalized_tree} in an exponential form.
Using Stirling's formula as $N \to \infty$ in the double scaling limit, $F^{\mathrm{tree}}$ can be approximated by
\begin{equation}
	F^{\mathrm{tree}} ( \epsilon ) \sim \frac{\epsilon}{2} \ln \epsilon - \frac{\epsilon}{2} - \frac{\epsilon}{2} \ln 8 \, .
\label{eq:F_tree}
\end{equation}
This tree-level contribution to the holy grail function is illustrated in Fig.~\ref{fig:F_tree}.

The observation that the amplitude $\mathcal{A}_N$ seems to take an exponential form in the large $N$ regime is not solely a mathematical statement about its structure.
It might also have physical consequences.
To begin with, there are two distinct points of $F^{\mathrm{tree}}$ which are of interest for our work -- the global minimum at $\epsilon_{\mathrm{min}} = 8$ and the root at $\epsilon_0 = 8e$ (cf.~Fig.~\ref{fig:F_tree}).
The crucial observation is that $F^{\mathrm{tree}}$ at $\epsilon_0$ changes from negative to positive sign, i.e.~the amplitude $\mathcal{A}_N^{\mathrm{tree}}$ will diverge in the limit $N \to \infty$ for any $\epsilon > \epsilon_0$
\begin{equation}
\lim_{N \to \infty} \mathcal{A}_N^{\mathrm{tree}} = \infty \, .
\end{equation}
This is unphysical since $\langle N|x|0\rangle>1/(2E_{N})$ is incompatible with the commutation relation $[\hat{x},\hat{p}]=i$~\cite{Brown:1992ay}.

This raises the question of how the behaviour of $F$ is changed when we include corrections to the tree-level result. 
Indeed, most interesting is the overall sign of $F$ for any $\epsilon$.
Consequently, our aim is to compute $F$ explicitly in the regime $N \to \infty$ with $\epsilon$ fixed.

In practice we will discover that to calculate these corrections we also need to go beyond leading order in $1/N$.
We will find that more generally we can reproduce the full perturbative series by writing\footnote{In principle one could be tempted to write $\mathcal{A}_N^{\mathrm{tree}}=\exp\left\{\frac{1}{\lambda}F^{\rm tree}(\epsilon,N)\right\}$ with $F^{\rm tree}(\epsilon,N)=F^{\rm tree}_{0}(\epsilon)+\frac{1}{N}F^{\rm tree}_{1}(\epsilon)+\ldots$.~However, using Stirling's formula for the factorial to higher orders we see that to write it in this form we have to factor out a $\lambda^{3/4}$. Moreover, one should also be mindful of the fact that Stirling's series is only asymptotic, suggesting missing pieces of the type we discussed at the beginning of this section.},
\begin{equation}
\mathcal{A}_N = \mathcal{A}_N^{\mathrm{tree}} \exp \left( \frac{1}{\lambda}  F_\Sigma(\lambda,N)  \right) \, 
\end{equation}
where 
\begin{equation}
F_{\Sigma} \left( \epsilon, N \right)  = F_0 \left( \epsilon \right) + \frac{F_1 \left( \epsilon \right)}{N} + \frac{F_2 \left( \epsilon \right)}{N^2} + \ldots \,,
\label{eq:F_schema}
\end{equation}
and where the $F_i \left( \epsilon \right)$ are analytic functions in $\epsilon$.

The polynomial structure of $\mathcal{A}_\Sigma$ derived in \eqref{eq:amplitude_factorize} tightly constrains the possible coefficients and powers of $\epsilon$ present in $F_\Sigma$.
In fact, the functional form of $F_\Sigma$ is given by
\begin{equation}
F_\Sigma \( \epsilon, N \) = \sum_{i,j=0}^{\infty} c_{ij} \frac{\epsilon^{i-j+2}}{N^j}  \enspace \mathrm{with} \enspace c_{ij} = 0 \; \quad \forall j > \frac{i+2}{2}
\label{eq:F_structure}
\end{equation}
because then a series expansion yields
\begin{equation}
\exp \( \frac{1}{\lambda} F_\Sigma \) = \sum_{k=0}^{\infty} \frac{\lambda^{-k}}{k!} \( c_{ij} \frac{1}{N^j} \epsilon^{i-j+2} \)^k = e^{c_{01}} \left[ 1 + \lambda \( c_{00} N^2 + c_{11} N + c_{22} \) + \mathcal{O} \( \lambda^2 \) \right]
\end{equation}
where the sum over $i$ and $j$ is understood.
This procedure gives a structure that could be described as a \textit{triangular expansion} of $F$ and can be schematically written as
\begin{equation}
F_\Sigma ( \epsilon, N ) \simeq
\begin{Bmatrix}
c_{00} \epsilon^2 & c_{01} \frac{1}{N} \epsilon^1 & & & \\
c_{10} \epsilon^3 & c_{11} \frac{1}{N} \epsilon^2 & & & \\
c_{20} \epsilon^4 & c_{21} \frac{1}{N} \epsilon^3 & c_{22} \frac{1}{N^2} \epsilon^2 & & \\
c_{30} \epsilon^5 & c_{31} \frac{1}{N} \epsilon^4 & c_{32} \frac{1}{N^2} \epsilon^3 & & \\
c_{40} \epsilon^6 & c_{41} \frac{1}{N} \epsilon^5 & c_{42} \frac{1}{N^2} \epsilon^4 & c_{43} \frac{1}{N^3} \epsilon^3 & \\
\vdots & & & & \ddots
\end{Bmatrix} \, .
\label{eq:F_triangle}
\end{equation}
The $i$-th column in fact corresponds to the terms of $F_i ( \epsilon )$ defined in \eqref{eq:F_schema}, for instance
\begin{equation}
F_0 \left( \epsilon \right) = c_{00} \epsilon^2 + c_{10} \epsilon^3 + c_{20} \epsilon^4 + \mathcal{O}\left( \epsilon^5 \right) \, .
\end{equation}

In general, the coefficients $c_{ij}$ can be matched to $\mathcal{A}_\Sigma$ by expanding the exponential and determining the missing coefficients at each order.
Using this matching $F(\epsilon)$ can be determined to arbitrary order in $\epsilon$ and $1/N$. In practice the computational effort increases rapidly.
For instance, the holy grail function corresponding to the factorized amplitude \eqref{eq:amplitude_factorize} is given by
\begin{equation}
F_{\Sigma} \left( \epsilon, N \right) =  - \frac{17}{16} \epsilon^2 + \frac{125}{64} \epsilon^3 + \mathcal{O} \left( \epsilon^4 \right) + \frac{1}{N} \left[ -\frac{5}{16} \epsilon^2 + \frac{99}{128} \epsilon^3 + \mathcal{O} \left( \epsilon^4 \right) \right] + \mathcal{O} \left( \frac{1}{N^2} \right) \, .
\label{eq:F_example}
\end{equation}

Remarkably, this method of reconstructing the holy grail function allows us to translate a series expansion of $\mathcal{A}_N$ in powers of $\lambda N^2$ into a series expansion of $F ( \epsilon, N )$ in powers of $\epsilon = \lambda N$.
Note that for the latter to be small is a much less restrictive statement.
Moreover the exact correspondence is a very powerful observation, because a finite number of coefficients $c_{ij}$ in $F$ will generate infinitely many terms of the amplitude. 
As discussed at the beginning of this section it is non-trivial that \textit{all} coefficients of perturbation theory can be recovered exactly.
We have checked that this is true to very high order.
For instance, we have verified that the first three non-trivial terms of $F_\Sigma$,
\begin{equation}
	F_\Sigma = - \frac{17}{16} \epsilon^2 + \frac{125}{64} \epsilon^3 - \frac{5}{16} \frac{\epsilon^2}{N} + \dots \, ,
\end{equation}
reproduce the first subleading corrections $\left( \lambda N^2 \right)^{k} / N$ of $\mathcal{A}_N$ up to order $k=15$.

\begin{figure}[t]
\centering
\includegraphics[width=0.5\textwidth]{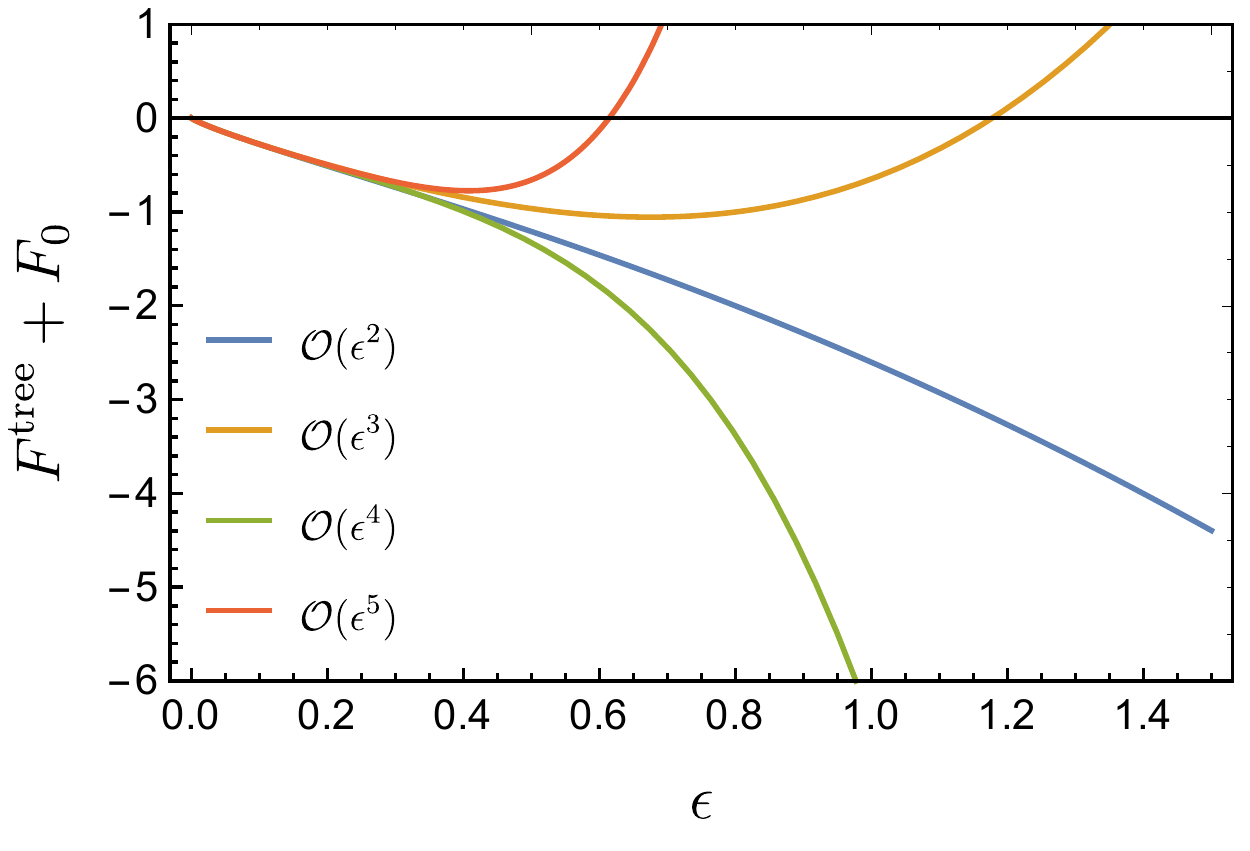}
\caption{Holy grail function $F = F^{\mathrm{tree}} + F_0$ in the double scaling limit $N\to\infty$, $\epsilon=\lambda N$ fixed, neglecting corrections of order $1/N$. The label denotes the highest order of $\epsilon$ that is included. The asymptotic behaviour for large $\epsilon$ is governed by the maximum order included in the series expansion, indicating that we have to apply resummation techniques.}
\label{fig:F_lead}
\end{figure}

\bigskip
As we have already mentioned all the important information about the $N \to \infty$ asymptotics of $\mathcal{A}_N$ is contained in
\begin{equation}
F_0 ( \epsilon ) = - \frac{17}{16} \epsilon^2 + \frac{125}{64} \epsilon^3 - \frac{17815}{3072} \epsilon^4 + \frac{87549}{4096} \epsilon^5 + \mathcal{O} \left( \epsilon^6 \right) \, .
\end{equation}
The holy grail function $F$ with the leading order corrections is shown in Fig.~\ref{fig:F_lead}.

We observe that $F_0$ is given by an alternating sum with monotonically growing coefficients.
Hence, naively the large $\epsilon$ asymptotics will be governed by the leading coefficient of the highest power in $\epsilon$.
Accordingly we cannot simply read off the value of $F$ for large $\epsilon$, because it depends on the truncation of the series expansion.
In order to still be able to extract a better estimate for $F$ we have to resum the perturbative series.
Since by construction we only know a finite (but still arbitrary) number of terms contained in $F$, we make use of a \textit{Pad\'e} approximation.

\begin{figure}[t]
\centering
\includegraphics[width=0.5\textwidth]{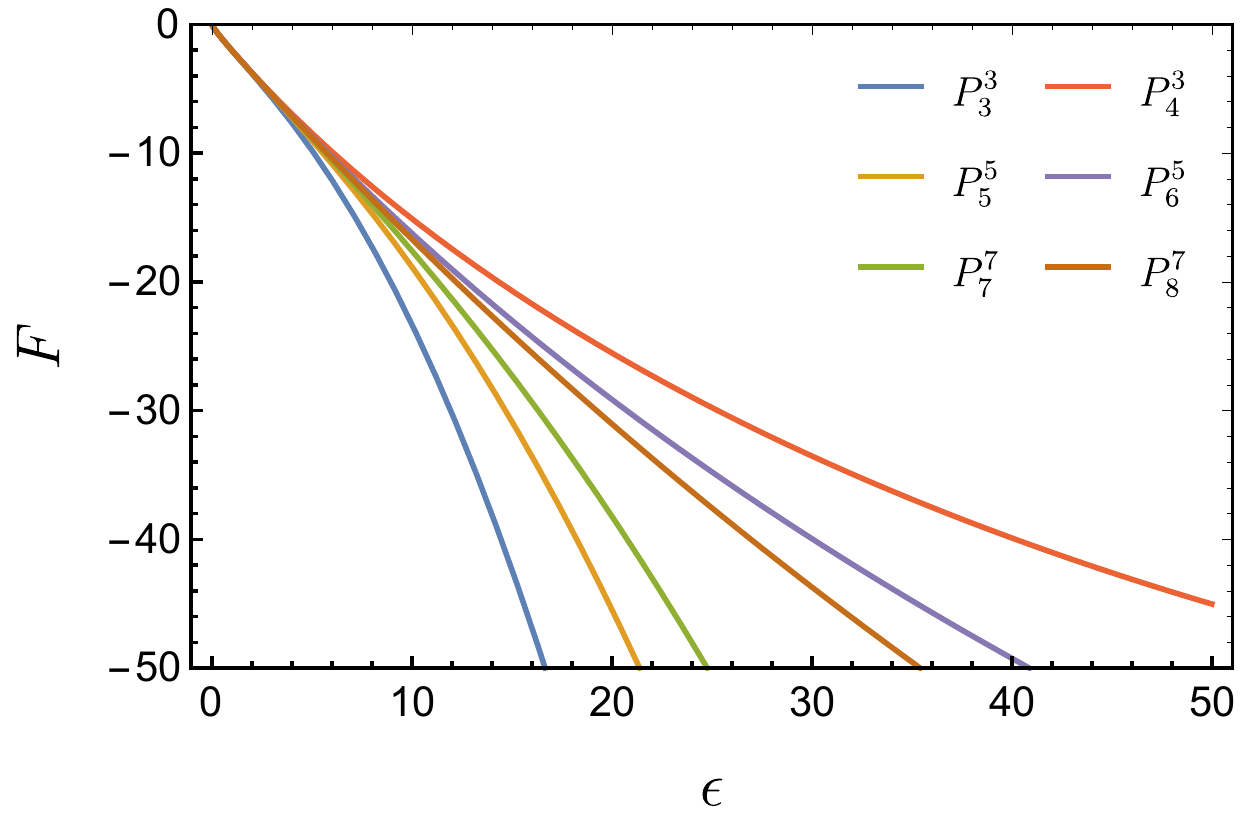}
\caption{Diagonal sequence of Pad\'e approximants of the holy grail function $F$  in the double scaling limit $N\to\infty$, $\epsilon=\lambda N$ fixed and at leading order in $1/N$, i.e.~$F=F^{\mathrm{tree}} + F_0$.}
\label{fig:F_pade_tot}
\end{figure}

To keep notation simple we use the standard symbols $P_m^n$ for the Pad\'e approximants defined by
\begin{equation}
P_m^n (\epsilon) \equiv \frac{\sum_{i=0}^n a_i \epsilon^i}{\sum_{j=0}^m b_j \epsilon^j}
\end{equation}
where one conventionally chooses $b_0 = 1$ without loss of generality.
As a standard technique of asymptotic analysis the idea is to consider the diagonal sequence $P_n^n$ and $P_{n+1}^n$.
Taylor expanding both and matching their coefficients $a_i$ and $b_j$ to the coefficients in $F$ yields the corresponding Pad\'e approximants at a given order $n$.
The first few approximants of that sequence are shown in Fig.~\ref{fig:F_pade_tot}.

\begin{figure}[t]
\centering
\begin{subfigure}{0.46\textwidth}
\includegraphics[width=\textwidth]{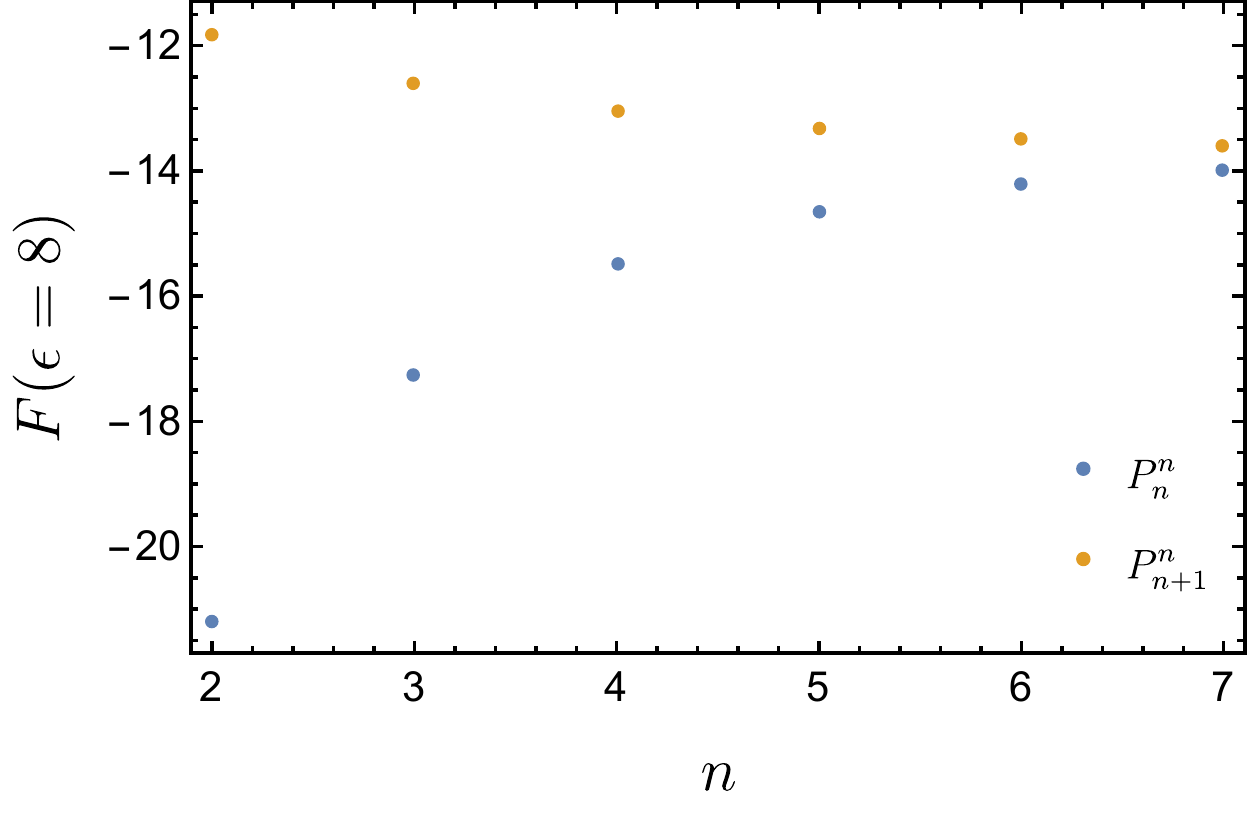}
\end{subfigure}
\hspace*{0.3cm}
\begin{subfigure}{0.46\textwidth}
\includegraphics[width=\textwidth]{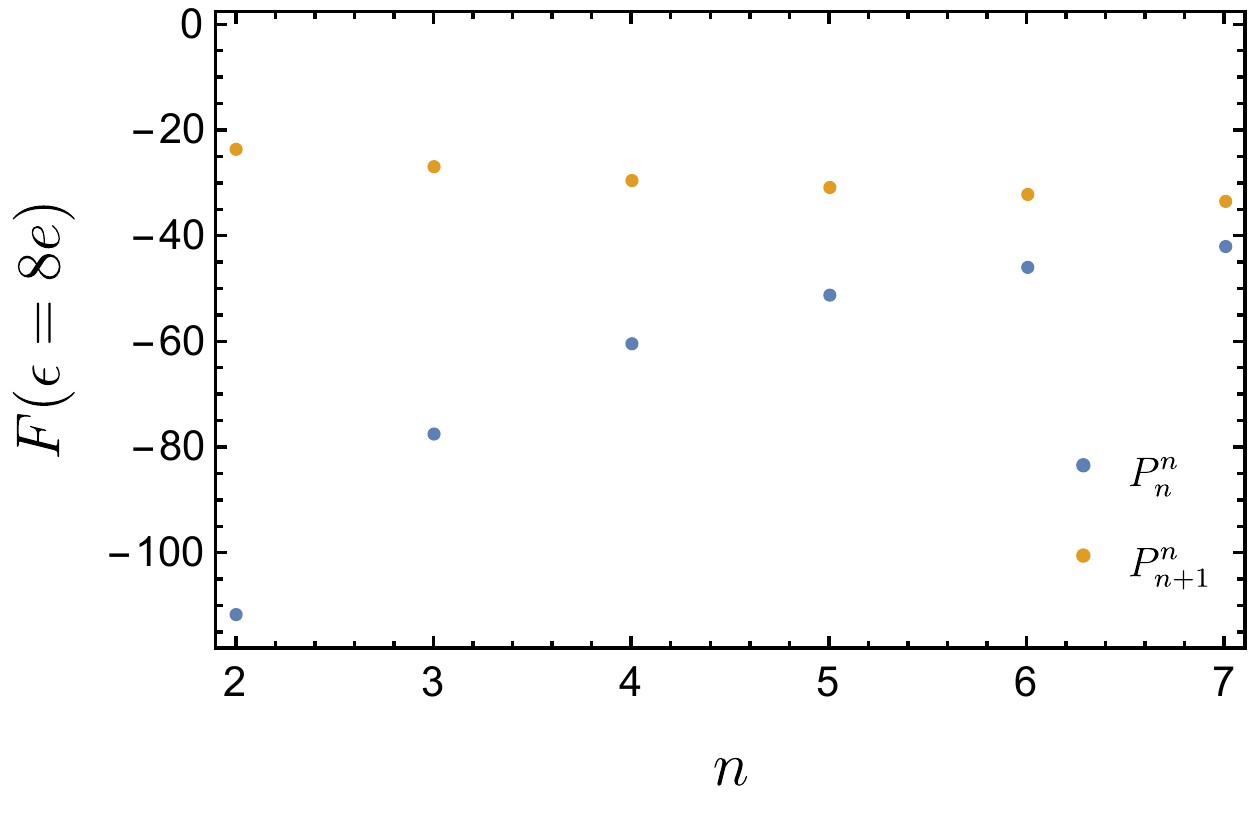}
\end{subfigure}
\caption{Pad\'e approximants of $F$ evaluated at the minimum $\epsilon = 8$ (left) and the root $\epsilon = 8e$ (right) of $F^{\mathrm{tree}}$. We use the limit $N\to\infty$, $\epsilon=\lambda N$ fixed and neglect corrections of order $1/N$.}
\label{fig:F_pade_min_root}
\end{figure}

One can clearly see (note the different scale in $\epsilon$) that resumming the holy grail function $F$ via a Pad\'e approximation drastically enhances the predictivity for large $\epsilon$.
The true value of the holy grail function $F$ for any value of $\epsilon$ is bounded from above and below by the Pad\'e approximants  $P_{n+1}^n ( \epsilon )$ and $P_{n}^n ( \epsilon )$.
When considering Pad\'e approximants $P_{n+1}^n$ of higher order the minima and roots are shifted towards larger $\epsilon$.
Additionally, the approximants $P_n^n$ are monotonically decreasing and do not exhibit any minima or roots at all.
Combining all these observations gives good evidence that the holy grail function $F$ remains negative for any value of $\epsilon$ in the limit $N \to \infty$, i.e.~the corresponding transition amplitude $\mathcal{A}_N$ does not diverge, but remains finite in that limit instead.
This observation is also explicitly supported by Fig.~\ref{fig:F_pade_min_root}, where the approximants are evaluated at the minimum $\epsilon = 8$ and root $\epsilon = 8e$ of the tree-level result.
They exhibit a nice convergence to a value of $F$, which is still negative at $\epsilon = 8e$.

\begin{figure}[t]
\centering
\includegraphics[width=0.5\textwidth]{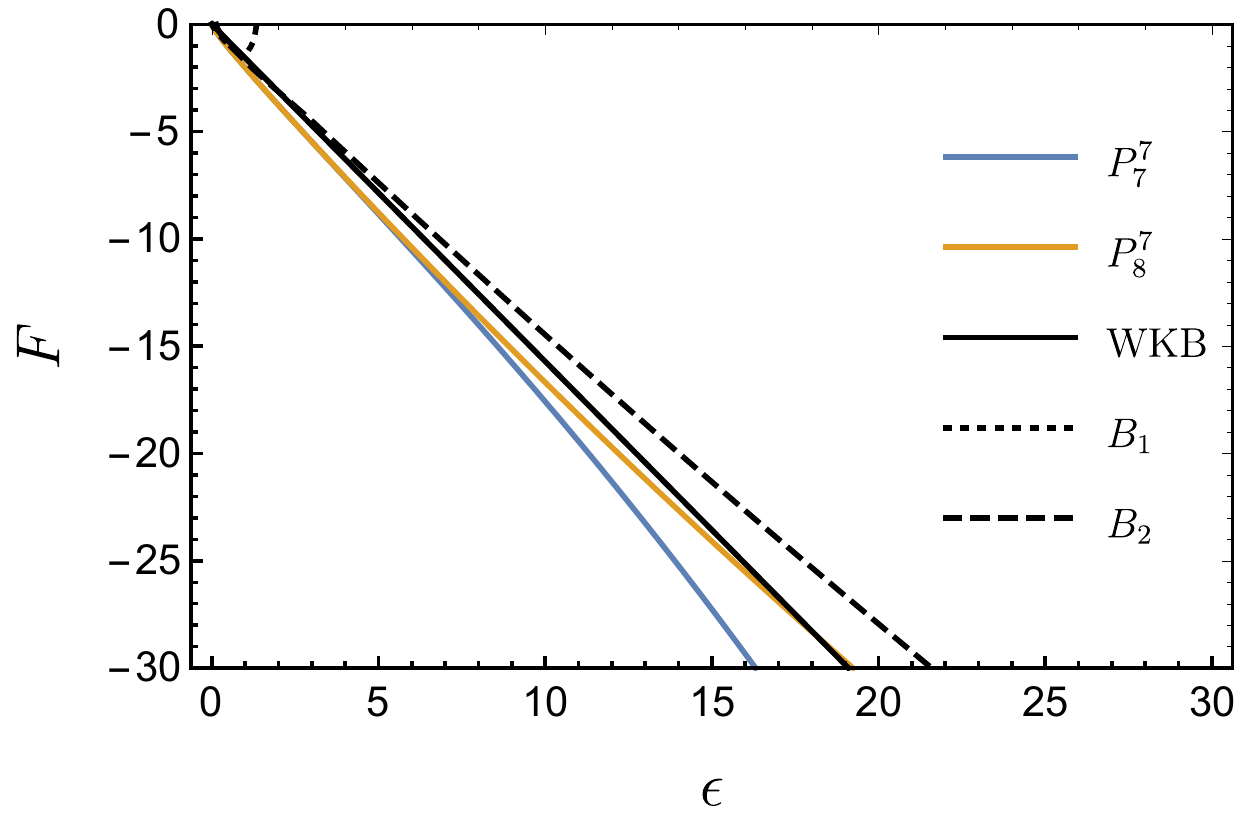}
\caption{Diagonal sequence of the highest Pad\'e approximants of the holy grail function $F$ at leading order in $1/N$ compared to existing results. $B_1$ and $B_2$ are rigorous bounds corresponding to the regimes $\epsilon \ll 1$ and $\epsilon \gg 1$ respectively \cite{Bachas:1991fd}. The label \textit{WKB} corresponds to a result obtained using complex WKB methods \cite{Cornwall:1993rh}. The Pad\'e resummed holy grail function appears to be consistent with both.}
\label{fig:F_pade_bounds}
\end{figure}

Furthermore, the Pad\'e resummed holy grail function $F$ that we systematically computed in this section is also consistent with two existing results on $\amp{N}$.
First of all in~\cite{Bachas:1991fd} Bachas gives a proof that $\lvert \amp{N} \rvert$ is non-trivially bounded from above.
If the amplitude is indeed of exponential form, this means that $F$ is also bounded and in particular negative for any $\epsilon$ in the double scaling limit $N\to\infty$, $\lambda N=const$.
More precisely in~\cite{Bachas:1991fd} explicit bounds on $F$ are given for $\epsilon \ll 1$ and $\epsilon \gg 1$, respectively.
Both limits can be translated into our variables and are shown in Fig.~\ref{fig:F_pade_bounds} denoted by $B_1$ and $B_2$.

Another result on $\amp{N}$ is worked out in \cite{Cornwall:1991gx, Cornwall:1992ip, Cornwall:1993rh}.
Using complex WKB methods the authors derive an explicit scaling of $\amp{N}$ for large $N$.
In particular, for $\epsilon \gg 1$ they obtain \cite{Cornwall:1993rh}
\begin{equation}
	\amp{N} \sim \exp \left(- \frac{\pi}{2} N \right) \, .
\label{eq:amplitude_cornwall}
\end{equation}
Again this explicit scaling is translated into our variables and illustrated in Fig.~\ref{fig:F_pade_bounds} labelled \textit{WKB}.

Fig.~\ref{fig:F_pade_bounds} shows that our result of systematically computing and resumming $F$ is consistent with the results of other works, indicating that in the double scaling limit the vacuum transition amplitude $\amp{N}$ indeed fully resums to an exponential governed by the holy grail function $F$.

We conclude that for the quartic anharmonic oscillator in the symmetric phase suitably resummed perturbation theory alone might be sufficient to make sense of diverging transition amplitudes for large excitation numbers.

\section{Transition Amplitudes for Arbitrary States Involving General Local Operators}
\label{sec:arbitrarylocal}

So far we have only considered transitions from the vacuum to some excited state, $\amp{N}$.
However, the same techniques to compute vacuum transitions can be applied to transition amplitudes between arbitrary states involving general local operators, $\kamp{N}{k}{M}$ with $k \in \mathbb{N}$\footnote{Due to the $\mathbb{Z}_2$ symmetry of the Hamiltonian, this amplitude is only non-vanishing, if $N+k+M$ is even.}.

In the following, we will argue that in the double scaling limit $N \to \infty$ with $\lambda N$ fixed the transition amplitudes are independent of the power of the local operator to exponential accuracy.
More precisely we find,
\begin{equation}
	\kamp{N}{k}{M} \sim R_k(N, M) \frac{\amp{N}}{\amp{M}}
\label{eq:gamplitude_reduction_schematic}
\end{equation}
where $R_k$ grows at most as a power of the quantum numbers $N$ and $M$.

In order to do so we derive the general form of $\kamp{N}{k}{M}$ and compare it to the right hand side of \eqref{eq:gamplitude_reduction_schematic}.
The general idea is to reduce the power of the operator that we are considering by insertions of the identity,
\begin{equation}
	\kamp{N}{k}{M} = \sum_L \kamp{N}{k-1}{L} \amp[M]{L} \, .
\end{equation}

Let us begin by considering the amplitude involving only a linear position operator ($k=1$).
In terms of the perturbative ansatz it reads (cf.~Section \ref{sec:perturbation_theory})
\begin{equation}
	\amp[M]{N} = \sum_{n=0}^\infty \lambda^n t_n^{N,M} \enspace \mathrm{with} \enspace t_n^{N,M} = \sum_{p=0}^n \sum_{k=0}^{N+4p} \sum_{l=0}^{M+4(n-p)} B_{p, k}^N B_{n-p,l}^M \Gamma \( \frac{k+l+2}{2} \) \, .
\label{eq:gamplitude_recursion}
\end{equation}
After normalizing both states we can extract the tree-level amplitude,
\begin{equation}
	\amp[M]{N}_{\mathrm{tree}} = \sqrt{\binom{N}{M} + \binom{M}{N}} \frac{\sqrt{\lvert N - M \rvert!}}{ 2^{\lvert N - M \rvert}} \left( \frac{\lambda}{4} \right)^{\frac{\lvert N - M \rvert - 1}{2}} \, .
\label{eq:gtree_amplitude}
\end{equation}
By examining the exponent of the coupling one can observe that the interpretation in terms of a field theory amplitude is not at all obvious -- for instance by increasing $M$ we effectively decrease the number of couplings that is needed for that particular transition from $M$ to $N$.
Naively, in the Feynman language of perturbative QFT we would expect exactly the opposite, because the number of couplings corresponds to the number of vertices in a given diagram.
However, $\amp[M]{N}_{\mathrm{tree}}$ contains information not only on the fully connected amplitude but also about disconnected pieces. Accordingly some care needs to be taken to establish a direct correspondence between $\amp[M]{N}$ and its field theory analogue.

Nevertheless, these transition amplitudes exhibit some very interesting features.
In particular, the form of the tree-level contribution shows a certain form of crossing symmetry and allows for a complete factorization into distinct amplitudes.
We observe that
\begin{equation}
\amp[M]{N}_{\mathrm{tree}} = \sqrt{\binom{N}{M} + \binom{M}{N}} \bra \lvert N-M \rvert | \hat{x} | 0 \ket_{\mathrm{tree}} \, .
\end{equation}
Taking $N>M$ and writing \eqref{eq:gtree_amplitude} as
\begin{equation}
\amp[M]{N}_{\mathrm{tree}} = \frac{\sqrt{N!}}{\sqrt{M!}}\frac{2^{-N/2}}{2^{-M/2}} \left( \frac{\lambda}{4} \right)^{\frac{N-M-1}{2}}
\end{equation}
it is easy to see that the tree-level contribution is in fact nothing but a quotient of two tree-level amplitudes.
Therefore, we can write
\begin{equation}
	\amp[M]{N}_{\mathrm{tree}} = \sqrt{\frac{M+1}{2}} \frac{\amp[K]{N}_{\mathrm{tree}} }{\amp[K]{M+1}_{\mathrm{tree}} }
\label{eq:amplitudefactorexp_tree}
\end{equation}
for an arbitrary state $\vert K \ket$ with $K \leq M$ and $N+K$ odd.
We could for example choose $K=0$ for $N$ odd and $K=1$ for $N$ even.

In fact these properties of the amplitude are not limited to the tree-level part, but also carry over to the full amplitude as we will show below.

Using the same methods as described in Section~\ref{sec:holy_grail} the higher order corrections to the amplitude can be computed. They are given by
\begin{equation}
\frac{\amp[M]{N}}{\amp[M]{N}_{\mathrm{tree}}}= 1 + \frac{\lambda}{16} \left( -17N^2 - 5N + 17M^2 + 29M + 12 \right) + \frac{\lambda^2}{512} \left( 289N^4 + 289M^4 + \ldots \right) \, .
\label{eq:gamplitude_normalized}
\end{equation}
Again it is surprising that both $N$ and $M$ completely decouple, i.e.~the first mixed terms $N^x M^y$ show up at quadratic order of the coupling $\mathcal{O} \left( \lambda^2 N^2 M^2 \right)$.
However, this decoupling in turn makes it straightforward to rewrite the amplitude into a holy grail function satisfying
\begin{equation}
\amp[M]{N} = \amp[M]{N}_{\mathrm{tree}} \exp \left( \frac{1}{\lambda} \mathcal{F}_\Sigma (\lambda, N, M) \right)
\end{equation}
where $\mathcal{F}_\Sigma$ is given by
\begin{equation}
\mathcal{F}_\Sigma (\lambda, N, M) = \lambda^2 \left( -\frac{17}{16} N^2 - \frac{5}{16} N + \frac{17}{16} M^2 + \frac{29}{16}M + \frac{3}{4} \right) + \mathcal{O} \left( \lambda^3 \right) \, .
\end{equation}

Furthermore, the decoupling of $N$ and $M$ implies that the holy grail function can be written as a sum of two components
\begin{equation}
\mathcal{F}_\Sigma (\lambda, N, M) = F_\Sigma(\lambda, N) + \hat{F}_\Sigma(\lambda, M)
\end{equation}
where $F_\Sigma(\lambda, N)$ is already known from the vacuum amplitude $\amp{N}$ given in \eqref{eq:F_example}.
Intriguingly, even the additional part $\hat{F}_\Sigma$ of the holy grail function can be fully recovered from $\amp{N}$ by observing
\begin{equation}
\hat{F}_\Sigma (\lambda, M) = - F_\Sigma(-\lambda, -(M+1)) \, .
\label{eq:Fhat_reduce}
\end{equation}
Hence, knowing $\amp{N}$ is in principle sufficient to reconstruct $\amp[M]{N}$ for arbitrary $M$.

The observation \eqref{eq:Fhat_reduce} allows us to extend the tree-level result \eqref{eq:amplitudefactorexp_tree} to the full amplitude.
In the leading $N$ and $M$ coefficients of the holy grail function (in the $1/N$-expansion), it can be rewritten as
\begin{equation}
	\amp[M]{N} \sim \sqrt{\frac{M+1}{2}} \frac{\amp[K]{N}}{\amp[K]{M+1}}
\label{eq:gamplitude_reduction}
\end{equation}
where $K \leq M$ is again an arbitrary quantum number, such that $N+K$ is odd.
Since we are neglecting terms of the form $\exp\left(1/N\right)$, it is valid to exponential accuracy.

The fact that transition amplitudes between arbitrary states reduce to a quotient of two vacuum transitions is not only surprising on its own, but is in fact a key to computing transitions for polynomials of local operators between arbitrary states
\begin{equation}
\bra N \lvert P \left(\hat{x}\right) \rvert M \ket = \sum_q a_q \kamp{N}{q}{M} \, .
\end{equation}
By generalizing the previous result \eqref{eq:gamplitude_reduction} to arbitrary powers of $\hat{x}$ we argue that the matrix element for any power of local operators reduces to vacuum transitions at leading order in $1/N$.
More precisely for $N>M$ the general claim for an arbitrary power of local operators is
\begin{equation}
	\kamp{N}{q}{M} \sim c_q \left( (N+1)^{\frac{3}{2}} - (M+1)^{\frac{3}{2}} \right)^{q-1} \frac{\amp[K_N]{N}}{\amp[K_M]{M+1}}
\label{eq:gamplitude_landau}
\end{equation}
where $c_q \in \mathbb{R}$ is a positive constant and $N+M+q$ and $M+K_M$ have to be even while $N+K_N$ is odd.
Again, this is due to the additional symmetry condition imposed by the $\mathbb{Z}_2$ symmetry of the Hamiltonian.
Since we demand $K_N < N$ and $K_M < M+1$, it is convenient to choose $K_{N,M} =0,1$, depending on the parity of the $N$-th and $M$-th level.

In the remaining part of this section we want to argue that \eqref{eq:gamplitude_landau} holds by using induction in $q$.
However, because of parity we have to consider even and odd $q$ separately.
For instance let us show \eqref{eq:gamplitude_landau} for even $N,M$ and $q$ explicitly.
In this case the claim, with $q=2p$ reads,
\begin{equation}
	\kamp{N}{2p}{M} \sim c_{2p} \left( (N+1)^{\frac{3}{2}} - (M+1)^{\frac{3}{2}} \right)^{2p-1} \frac{\amp[1]{N}}{\amp{M+1}} \, .
\label{eq:gamplitude_landau_qeven}
\end{equation}
The first non-trivial case is $p=1$ for which a full derivation can be found in appendix \ref{app:quadposoperator}. However, we just state the result here,
\begin{equation}
	\kamp{N}{2}{M} \sim \frac{(N+1)^{\frac{3}{2}} - (M+1)^{\frac{3}{2}}}{3\sqrt{2}} \sqrt{\frac{M+1}{2}} \frac{\amp[1]{N}}{\amp{M+1}} \, .
\label{eq:gamplitude_landau_qeven_1}
\end{equation}

Having established the first non-trivial case, we can proceed by considering $p \to p+1$ and inserting the identity operator, i.e.
\begin{equation}
	\kamp{N}{2(p+1)}{M} = \sum_{L=0}^{\infty} \kamp{N}{2p}{L} \kamp{L}{2}{M} \, .
\end{equation}
Splitting the sum into three different contributions depending on the parameter ranges of $L$ and applying the induction hypothesis \eqref{eq:gamplitude_landau_qeven} together with the initial result \eqref{eq:gamplitude_landau_qeven_1}, we obtain
\begin{equation}
	\kamp{N}{2(p+1)}{M} \sim c_{2p} \left( S_1 + S_2 + S_3 \right)
\end{equation}
where we have
\begin{align}
	S_1(N) &\equiv \amp[1]{N} \amp[1]{M} 
	\\\nonumber
	&\quad\quad\quad\quad\quad\quad\times\sum_{L=0}^M \frac{L+1}{2} \left[(N+1)^{\frac{3}{2}} - (L+1)^{\frac{3}{2}} \right]^{2p-1} \frac{(M+1)^{\frac{3}{2}} - (L+1)^{\frac{3}{2}}}{3\sqrt{2}} \frac{1}{\amp{L+1}^2} \\
	S_2(N) &\equiv \sqrt{\frac{M+1}{2}} \frac{\amp[1]{N}}{\amp{M+1}}
	\\\nonumber
	&\quad\quad\quad\quad\quad\times \sum_{L=M}^N \sqrt{\frac{L+1}{2}} \left[(N+1)^{\frac{3}{2}} - (L+1)^{\frac{3}{2}} \right]^{2p-1} \frac{(L+1)^{\frac{3}{2}} - (M+1)^{\frac{3}{2}}}{3 \sqrt{2}} \frac{\amp[1]{L}}{\amp{L+1}} \\
	S_3(N) &\equiv \sqrt{\frac{(N+1)(M+1)}{2}}\frac{1}{\amp{N+1} \amp{M+1}} 
	\\\nonumber
	&\quad\quad\quad\quad\quad\quad\quad\quad\quad\quad\times\sum_{L=N}^{\infty} \left[(L+1)^{\frac{3}{2}} - (N+1)^{\frac{3}{2}} \right]^{2p-1} \frac{(L+1)^{\frac{3}{2}} - (M+1)^{\frac{3}{2}}}{3\sqrt{2}} \amp[1]{L}^2 \, .
\end{align}
That is, we have three contributions $S_1,S_2$ and $S_3$ to the leading $N$ behaviour of the amplitude.
Similar to the case $p=1$ (cf.~appendix \ref{app:quadposoperator}) these can be analysed independently with respect to their asymptotics for large $N$.
\begin{enumerate}[(i)]
\item
The first contribution $S_1$ contains a sum that is only explicitly dependent on $N$.
That is, we can just pick the term with the highest power of $N$ as the dominant contribution to $S_1$.
This gives
\begin{equation}
	S_1(N) \sim (N+1)^{\frac{3}{2}(2p-1)} \amp[1]{N} \, .
\end{equation}

\item
In contrast to $S_1$ the second contribution $S_2$ involves a sum that depends on $N$ both explicitly and implicitly (since it appears as a boundary term).
However, by observing that $\amp[1]{L} / \amp{L+1} \sim \mathcal{O}(1)$ we can evaluate the sum explicitly by rewriting it as an integral.
Then considering that the sum contains only even $L$, this leads to
\begin{equation}
	S_2(N) \sim \frac{\left[(N+1)^{\frac{3}{2}} - (M+1)^{\frac{3}{2}} \right]^{2p+1}}{\left(3 \sqrt{2} \right)^2 2p (2p+1)} \amp[1]{N} \, .
\end{equation}

\item
Unfortunately, the last contribution $S_3$ cannot be carried out in full detail, because the sum not only depends on $N$ but also on the form of $\amp[1]{L}$.
However, since we are only interested in the parametric dependence on $N$, we can use that we expect the amplitudes to be of exponential form (cf.~\eqref{eq:amplitude_cornwall})
\begin{equation}
	\amp[0,1]{N} \sim e^{-cN}
\end{equation}
where $c$ is a positive constant.
Using this parametric ansatz for $\amp[1]{L}$ we can establish an upper bound for the sum contained in $S_3$ by writing
\begin{equation}
	\sum_{L=N}^{\infty} \left[ (L+1)^{\frac{3}{2}} \right]^{3p} \amp[1]{L}^2 \sim \int_{L=N/2}^{\infty} (2L+1)^{3p} e^{-4cL} = \frac{1}{2} e^{2c} (N+1)^{3p+1} E_{-3p}\left[2c(N+1) \right]
\end{equation}
where $E_n(z)$ denotes the exponential integral function and we used that the summation contains only even $L$.
Using the asymptotic expansion of $E_n(z)$
\begin{equation}
E_n(z) \sim \frac{e^{-z}}{z} \left[ 1 - \frac{n}{z} + \mathcal{O}\left(\frac{n^2}{z^2} \right) \right] \qquad (z \to \infty)
\end{equation}
we can infer that the dominant terms of $S_3$ are at most of order 
\begin{equation}
	S_3(N) \lesssim (N+1)^{\frac{3}{2} \left( 2p + \frac{1}{3} \right)} \sqrt{\frac{M+1}{2}} \frac{\amp{N+1}}{\amp{M+1}} \, .
\end{equation}
\end{enumerate}

Comparing the asymptotic terms for large $N$ of $S_1, S_2$ and $S_3$, we can conclude that the dominant contribution is given by $S_2$.
Finally, we obtain
\begin{equation}
	\kamp{N}{2(p+1)}{M} \sim c_{2(p+1)} \left[ (N+1)^{\frac{3}{2}} - (M+1)^{\frac{3}{2}} \right]^{2(p+1)-1} \sqrt{\frac{M+1}{2}} \frac{\amp[1]{N}}{\amp{M+1}}
\end{equation}
which is exactly the induction hypothesis for $p+1$.

Even though we will not give the full argument, it is straightforward to do the same computation for odd $N$ and $M$ and also for odd $q$ (with $N$ and $M$ of different parity) by doing the same manipulations to the initial amplitude.

In summary, we find that to exponential accuracy for any power of the local operator the corresponding transition amplitude is equal to the linear one,
\begin{equation}
	\kamp{N}{q}{M} \sim c_q \left( (N+1)^{\frac{3}{2}} - (M+1)^{\frac{3}{2}} \right)^{q-1} \frac{\amp[K_N]{N}}{\amp[K_M]{M+1}}
\end{equation}
where $c_q \in \mathbb{R}$ is a positive constant and $N+M+q$ and $M+K_M$ have to be even while $N+K_N$ is odd.

As a particular example this implies that
\begin{equation}
	\kamp{N}{q}{0} \sim c_q (N+1)^{\frac{3}{2} (q-1)} \amp{N} \, .
\label{eq:kamp_independence}
\end{equation}
This suggests that to exponential accuracy and in the double scaling limit $\amp{N}$ contains all the information on $\kamp{N}{q}{0}$.
Consequently, by comparison to previous results on the vacuum transition $\amp{N}$ we can conclude that $\kamp{N}{q}{0}$ also remains finite as $N \to \infty$.

Finally, our result supports the assumption that, to exponential accuracy, the amplitude in question is independent of the precise form of the local operator, e.g.
\begin{equation}
	\langle N|\phi|0\rangle\sim \langle N|\phi^2 |0\rangle \, .
\label{eq:independence}
\end{equation}
This is one of the ingredients in semiclassical calculations.
However, some caution is needed in its direct application because the semiclassical methods of~\cite{Son:1995wz,Khoze:2017ifq} make use of an exponential operator $\exp(j\hat{\phi})$.
For finite $j$ Eq.~\eqref{eq:kamp_independence} is not sufficient to guarantee that there are no exponential prefactors.
Hence, the limit $j\to 0$ has to be taken with care.

\section{Conclusions}
\label{sec:conclusions}
High multiplicity amplitudes in scalar quantum field theories have recently attracted renewed attention~\cite{Jaeckel:2014lya,Khoze:2017tjt,Khoze:2017lft, Khoze:2017uga, Khoze:2018bwa}.
Perturbative as well as semiclassical calculations indicate a potential growth of these amplitudes~\cite{Cornwall:1990hh, Goldberg:1990qk, Brown:1992ay, Voloshin:1992mz, Argyres:1992np, Smith:1992kz, Smith:1992rq,Son:1995wz,Khoze:2017ifq}.
This raises questions about the consistency of the calculation and perhaps even the theory itself~\cite{Jaeckel:2014lya}, but it could also lead to an interesting solution of the hierarchy problem~\cite{Khoze:2017tjt}.

To shed light on this we investigate the anharmonic quantum mechanical oscillator which is the analogue of $\phi^4$-theory. In contrast to most studies of this system (notable exceptions are~\cite{Bachas:1991fd,Bachas:1992dw,Brown:1992ay,Cornwall:1991gx,Cornwall:1992ip,Cornwall:1993rh,Khlebnikov:1994dc}) we focus on the transition amplitudes $\langle N|\hat{x}|0\rangle$ that are the analogue to the high multiplicity amplitudes.
Our results for the system with a single minimum can be summarized as follows:
\begin{itemize}
\item{} The perturbative series can be reproduced exactly by,
\begin{equation}
	\amp{N} = \amp{N}_{\rm tree}\exp \left( \frac{1}{\lambda} F_{\Sigma} \left( \lambda, N \right) \right),
\end{equation}
where 
\begin{equation}
	F_{\Sigma} \left( \epsilon, N \right) = F_0 \left( \epsilon \right) + \frac{F_1 \left( \epsilon \right)}{N} + \frac{F_2 \left( \epsilon \right)}{N^2} + \mathcal{O} \left( \frac{1}{N^3} \right).
\end{equation}
We have explicitly checked this to a significant order in perturbation theory providing additional evidence
for this form conjectured in~\cite{Voloshin:1992nu,Khlebnikov:1992af,Libanov:1994ug,Libanov:1995gh,Bezrukov:1995qh,Libanov:1996vq,Son:1995wz}. We have clarified that in the anharmonic oscillator with quartic coupling this seems to be an exact correspondence 
order by order in $\epsilon$ and $1/N$. (We will consider more general potentials in future work~\cite{inpreparation}.)
\item{} In the double scaling limit $N \to \infty$ and $\lambda \to 0$ with $\epsilon=\lambda N$ fixed the asymptotic behaviour of $\amp{N}$ is governed by $F^{\mathrm{tree}} \left( \epsilon \right) + F_0 \left( \epsilon \right)$. 
Using Pad\'e resummation the perturbative behaviour is significantly improved and we find strong indications that
\begin{equation}
F^{\mathrm{tree}} \left( \epsilon \right) + F_0 \left( \epsilon \right)<0 \quad\forall\, \epsilon \, .
\end{equation}
This avoids problems with unitarity and existing bounds from~\cite{Bachas:1991fd, Brown:1992ay}.
\item{} These results can be generalized to a larger class of amplitudes,
\begin{equation}
	\amp[M]{N} = \amp[M]{N}_{\mathrm{tree}} \exp \left( \frac{1}{\lambda} {\mathcal F}_\Sigma (\lambda, N, M) \right)
\end{equation}
where $\mathcal{F}_\Sigma$ is schematically given by
\begin{equation}
	\mathcal{F}_\Sigma (\lambda, N, M) = F_\Sigma(\lambda, N) - F_\Sigma(-\lambda, -(M+1)) \, .
\end{equation}
In particular the amplitude factorizes into two distinct pieces.

\item{} We also confirm the conjecture that to exponential accuracy the amplitude is independent of the precise form
of the inserted local operator\footnote{As long as the operator in question is polynomial in the field operator.}, a feature already mentioned in~\cite{Landau:1991wop}.
We obtain
\begin{equation}
	\kamp{N}{q}{M} \sim c_q \left( (N+1)^{\frac{3}{2}} - (M+1)^{\frac{3}{2}} \right)^{q-1} \frac{\amp[K_N]{N}}{\amp[K_M]{M+1}} \, .
\end{equation}
\end{itemize}

In our toy model, our results indicate that suitable resummation of the perturbative coefficients can prevent the growth of high multiplicity amplitudes and preserves unitarity. This indicates a possible path towards how such a strong growth at high multiplicities could be resolved also in the case of the Standard Model Higgs.
However, we note that there are two crucial differences between our toy model and the Higgs.
The first is that the Higgs is a spontaneously broken theory and already in quantum mechanics this leads to the relevance of non-perturbative instantonic configurations.
Secondly, the Higgs is a four-dimensional quantum field theory.
This allows for the final states to disperse into space after the interaction, something that is not possible in quantum mechanics.
Both of these merit further investigation.

\section*{Acknowledgements}
We would like to thank Valya Khoze, Michael Spannowsky and Zsolt Szabo for many interesting and helpful discussions. SS gratefully acknowledges financial support by the Heidelberg Graduate School of Fundamental Physics.

\appendix

\section{Transition Amplitudes Involving Quadratic Position Operators}
\label{app:quadposoperator}

As the simplest example of higher power local operators let us consider a quadratic transition $\kamp{N}{2}{M}$ with $N > M$.
Due to the $\mathbb{Z}_2$ symmetry of the Hamiltonian the states $N$ and $M$ have to be of the same parity in order to obtain a non-vanishing result.
Consequently there are two cases (even and odd parity) that we have to consider.
These are, however, very similar in the computation.
We will therefore demonstrate the relation for $N,M$ odd and just give the result for $N,M$ even.

Let us begin by inserting an identity operator into $\kamp{N}{2}{M}$,
\begin{equation}
	\kamp{N}{2}{M} = \sum_{L=0}^\infty \amp[L]{N} \amp[M]{L}
\end{equation}
where $L$ has to be even in this case.
This effectively reduces the power of the operator at the cost of introducing an infinite sum.
However, splitting the sum into three different pieces and using \eqref{eq:gamplitude_reduction} we can write
\begin{equation}
\begin{split}
	\kamp{N}{2}{M} \sim
	& \amp{N} \amp{M} \sum_{L=0}^M \frac{L+1}{2} \frac{1}{\amp{L+1}^2} \\
	& + \sqrt{\frac{M+1}{2}} \frac{\amp{N}}{\amp[1]{M+1}} \sum_{L=M}^N \sqrt{\frac{L+1}{2}} \frac{\amp[1]{L}}{\amp{L+1}} \\
	& +\sqrt{\frac{N+1}{2}} \sqrt{\frac{M+1}{2}} \frac{1}{\amp[1]{N+1}} \frac{1}{\amp[1]{M+1}} \sum_{L=N}^\infty \amp[1]{L}^2 \, .
\end{split}
\end{equation}

Since we are only interested in the leading terms in $N$ of $\kamp{N}{2}{M}$, we have to determine which of the three contributions is dominant for large $N$.
We will discuss them separately in order of appearence.
\begin{enumerate}[(i)]

\item
The first contribution
\begin{equation}
	S_1(N) \equiv \amp{N} \amp{M} \sum_{L=0}^M \frac{L+1}{2} \frac{1}{\amp{L+1}^2}
\end{equation}
is obvious to determine, since the sum does not involve any $N$ and thus we conclude for the leading terms of $S_1$
\begin{equation}
	S_1(N) \sim \amp{N} \, .
\end{equation}

\item
The second contribution
\begin{equation}
	S_2(N) \equiv \sqrt{\frac{M+1}{2}} \frac{\amp{N}}{\amp[1]{M+1}} \sum_{L=M}^N \sqrt{\frac{L+1}{2}} \frac{\amp[1]{L}}{\amp{L+1}}
\end{equation}
is a little bit more complicated to determine than $S_1$, because the sum now involves $N$ as a boundary term.
However, since we are mainly interested in the parametric dependence for large $N$ we can estimate the sum by using $\amp[1]{L}/\amp{L+1} \sim \mathcal{O}(1)$.
We can then carry it out explicitly by going to the continuum limit, i.e.~to an integral which reads
\begin{equation}
	\sum_{L=M}^N \sqrt{\frac{L+1}{2}} = \int_{\frac{M}{2}}^{\frac{N}{2}} dL \sqrt{L+\frac{1}{2}} = \frac{(N+1)^{\frac{3}{2}} - (M+1)^{\frac{3}{2}}}{3\sqrt{2}}
\end{equation}
where we used that the sum only includes terms with even $L$.
The dominant terms in $N$ then are
\begin{equation}
	S_2(N) \sim \left( (N+1)^{\frac{3}{2}} - (M+1)^{\frac{3}{2}} \right) \amp{N} \, .
\end{equation}

\item
The third contribution
\begin{equation}
	S_3(N) \equiv \sqrt{\frac{N+1}{2}} \sqrt{\frac{M+1}{2}} \frac{1}{\amp[1]{N+1}} \frac{1}{\amp[1]{M+1}} \sum_{L=N}^\infty \amp[1]{N}^2
\end{equation}
is the most involved one, because the sum involves $N$ not only as a boundary term but also the \textit{a priori} unknown amplitudes $\amp[1]{L}$.
However, if we think back to earlier results, we know that all amplitudes should parametrically be of exponential form (cf.~\eqref{eq:amplitude_cornwall}), i.e.
\begin{equation}
	\amp[0,1]{N} \sim e^{-cN}
\end{equation}
where $c>0$.
Similar to $S_2$ using this ansatz the sum can be continued to an integral, such that we obtain
\begin{equation}
	\sum_{L=N}^\infty \amp[1]{L} \sim e^{-2cN} \sim \amp{N}^2
\end{equation}
where we used the parametric exponential dependence of the amplitude twice.
Thus, the third contribution reads in the leading terms in $N$
\begin{equation}
	S_3(N) \sim \sqrt{\frac{N+1}{2}} \amp[1]{N+1} \, .
\end{equation}
\end{enumerate}

If we compare all three different contributions $S_1, S_2$ and $S_3$ for large $N$, we conclude that parametrically $S_2$ is the dominant contribution to $\kamp{N}{2}{M}$.
In summary, we can write
\begin{equation}
	\kamp{N}{2}{M} \sim \frac{(N+1)^{\frac{3}{2}} - (M+1)^{\frac{3}{2}}}{3\sqrt{2}} \sqrt{\frac{M+1}{2}} \frac{\amp{N}}{\amp[1]{M+1}}
\end{equation}
for $N,M$ odd and $N > M$.
In a similar way, the same result can also be obtained for $N, M$ even.
The only difference is that now the low lying states are exchanged because of parity conservation,
\begin{equation}
	\kamp{N}{2}{M} \sim \frac{(N+1)^{\frac{3}{2}} - (M+1)^{\frac{3}{2}}}{3\sqrt{2}} \sqrt{\frac{M+1}{2}} \frac{\amp[1]{N}}{\amp{M+1}} \, .
\end{equation}

\bibliographystyle{h-physrev}

\end{document}